\documentclass[12pt]{iopart}
\usepackage{graphicx}

\begin{document}

\title{Fermion contribution to the
static quantities of arbitrarily charged vector bosons}
\author{G. Tavares-Velasco  and J. J. Toscano}
\address{Facultad de Ciencias F\'\i sico Matem\' aticas, Benem\' erita Universidad
Aut\' onoma de Puebla, Apartado Postal 1152, Puebla, Pue., M\'
exico}

\begin{abstract}
We present an analysis of the one-loop contribution from left- and
right-handed fermions to the static electromagnetic properties of
an arbitrarily charged no self-conjugate vector boson $V$.
Particular emphasis is given to the case of a no self-conjugate
neutral boson $V^0$. Regardless the electric charge of the $V$
boson, a fermionic loop can induce the two CP-even form factors
but only one CP-odd. As a result the corresponding electric dipole
moment is directly proportional to the magnetic quadrupole moment.
The CP-odd form factor might be severely suppressed since it
requires the presence of both left- and right-handed fermions. The
behavior of the form factors is analyzed for several scenarios of
the fermion masses in the context of the decoupling theorem.
\end{abstract}

\submitto{\JPG}
\maketitle

\section{Introduction}
It is well known that charged particles with nonzero spin can
interact with the electromagnetic field not only via the monopole
but also through higher order multipoles. Though these particles
do receive contributions beyond the monopole at the tree level, it
is interesting to investigate those contributions arising from
quantum fluctuations as they can be sensitive to new physics
effects. In particular, the static electromagnetic properties of
charged vector bosons are parametrized by four loop-generated
electromagnetic form factors: two CP-even ones ($\Delta \kappa$
and $\Delta Q$) and two CP-odd ones ($\Delta \widetilde{\kappa}$
and $\Delta \widetilde{Q}$). The CP-even form factors define the
(anomalous) magnetic dipole and electric quadrupole moments of the
vector boson, whereas the CP-odd ones parametrize its electric
dipole and magnetic quadrupole moments. These form factors are
model dependent and can only be generated at the one-loop level or
higher orders. For instance, in the case of the charged $W$ gauge
boson of the standard model (SM), only the CP-even quantities are
generated at the one-loop level \cite{Bardeen}, whereas the CP-odd
ones can only be induced at higher orders \cite{Lopez}. Because of
their sensitivity to new physics effects, the electromagnetic
properties of the $W$ boson have been the subject of considerable
interest and have been studied in many extensions of the SM
\cite{WWg}. As far as neutral particles are concerned, by
time-reversal invariance they cannot have static electromagnetic
properties as long as they are characterized by self-conjugate
fields, in which case the particle coincides with the
antiparticle. This is true for instance for massive Majorana
neutrinos \cite{Majorana} and the neutral gauge boson of the SM
\cite{Z}, which only have off-shell electromagnetic properties.
The situation is different for neutral particles characterized by
no self-conjugate fields, in which case the particle does not
coincides with the antiparticle. For instance, massive Dirac
neutrinos \cite{Dirac} or complex neutral vector bosons
\cite{Nieves} do can have static electromagnetic properties. In
contrast with the case of charged particles, the electromagnetic
properties of neutral no self-conjugate particles can only arise
from quantum fluctuations, thereby being rather sensitive to
virtual effects from new particles.  Even though neutral
self-conjugate vector bosons are not predicted by the SM, they can
arise in some of its extensions \cite{Long} and so it is
interesting to study their static electromagnetic properties.

Although the $VV^\dag\gamma$ vertex can receive contributions from
fermion, boson and scalar loops, we will concentrate only on the
fermion contributions. The reason for this choice is that in most
of the renormalizable theories the boson or scalar particles
cannot generate any CP-odd form factor at the one-loop level
\cite{Burgess}, whereas both the CP-even and CP-odd form factors
do can be generated at this order by left- and right-handed
fermions. This means that, the CP-odd form factors will be
entirely determined by the fermion contribution in most of the
renormalizable theories. In addition, while the couplings of $V$
to gauge boson or scalar pairs would be more model-dependent, a
renormalizable coupling of $V$ to a fermion pair can be proposed
by minimal substitution.

This paper has been organized as follows. In Sec. \ref{cal} the
calculation of the amplitude is presented.  Sec. \ref{dis} is
devoted to analyze the behavior of the electromagnetic properties
of the $V$ boson in diverse scenarios of the fermion masses.
Emphasis is given to the properties of the form factors in the
context of the decoupling theorem \cite{Appelquist}. The
conclusions are presented in Sec. \ref{con}.

\section{The on-shell
$VV^\dag\gamma$ vertex} \label{cal} According to the notation
shown in figure 1, the most general on-shell $VV^{\dag}\gamma$
vertex can be written as \cite{Nieves,Hagiwara}
\begin{eqnarray}
\label{v} \Gamma_{\alpha \beta \mu}&=&ie\Bigg\{A_V[2p_\mu
g_{\alpha \beta} +4(q_\beta g_{\alpha \mu}-q_\alpha g_{\beta
\mu})]+2\Delta \kappa_V (q_\beta g_{\alpha \mu}-q_\alpha g_{\beta
\mu})\nonumber
\\
&&+\frac{4\Delta Q_V}{m^2_{V}}p_\mu q_\alpha q_\beta +2\Delta
\widetilde{\kappa}_V\epsilon_{\alpha \beta \mu
\lambda}q^\lambda+\frac{4\Delta \widetilde{Q}_V}{m^2_{V}}q_\beta
\epsilon_{\alpha \mu \lambda \rho}p^\lambda q^\rho\Bigg\}.
\end{eqnarray}
In the case of a neutral vector boson, the $A_V$ coefficient
vanishes at any order of perturbation theory since it violates
gauge invariance. The magnetic (electric) dipole moment $\mu_{V}$
($\widetilde {\mu}_{V}$) and the electric (magnetic) quadrupole
moment $Q_{V}$ ($\widetilde{Q}_{V}$) are given in terms of the
electromagnetic form factors as follows
\begin{eqnarray}
\mu_{V}&=&\frac{e}{2m_{V}}(2+\Delta \kappa_V), \\
Q_{V}&=&-\frac{e}{m^2_{V}}(1+\Delta \kappa_V+\Delta Q_V),\\
\widetilde{\mu}_{V}&=&\frac{e}{2m_{V}}\Delta \widetilde{\kappa}_V,\\
\widetilde{Q}_{V}&=&-\frac{e}{m^2_{V}}(\Delta
\widetilde{\kappa}_V+\Delta \widetilde{Q}_V).
\end{eqnarray}

\begin{figure}
\center\includegraphics[width=2.5in]{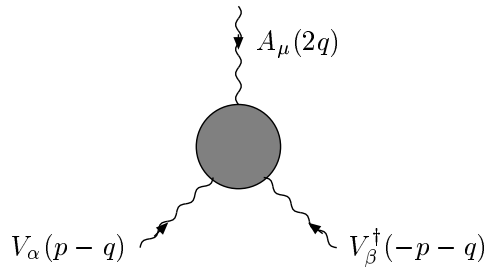} \caption{The
$VV^{\dag}\gamma$ vertex. The large dot represents the virtual
effects of left- and right-handed fermions.}
\end{figure}

As for the fermion contribution, it has been pointed out in Ref.
\cite{Burgess} that the only fermion interaction that can
contribute to the on-shell $VV^{\dag}\gamma$ vertex at the
one-loop level is the following fermion-gauge coupling:
\begin{equation}
{\cal L}=\bar{\psi}_1\gamma_\mu (g_LP_L+g_RP_R)\psi_2\,V^\mu +{\rm
H.c},
\end{equation}
where $g_L$ and $g_R$ are arbitrary complex parameters. We will
present a general calculation and then discuss some specific
scenarios, in particular that of a neutral no self-conjugate
vector boson. The contribution of the fermion pair ($\psi_1$,
$\psi_2$) to the $VV^\dag\gamma$ vertex is given by the two
Feynman diagrams shown in figure 2. We will denote by $Q_1$ and
$Q_2$ the charges of the fermions circulating in the loop. The
charge of the $V$ boson is thus $Q_V=Q_2-Q_1$. The respective
amplitude can be written down readily and the loop integral can be
solved via the Feynman parameter technique. Below we will present
separately the results for the CP-even and CP-odd form factors.

\begin{figure}
\center\includegraphics[width=4in]{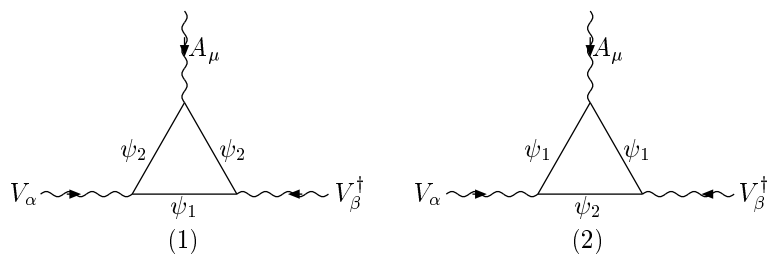} \caption{Feynman
diagrams contributing to the $VV^{\dag}\gamma$ vertex. $\psi_i$ is
a fermion with left- and right-handed couplings to the arbitrarily
charged $V$ boson.}
\end{figure}

\subsection{CP-even form factors}
Once the integration over the arbitrary momentum is done, the
following CP-even form factors are obtained
\begin{eqnarray}
\Delta
\kappa_V=\frac{Q_1}{16\pi^2}\left(\left(|g_L|^2+|g_R|^2\right)2\,{\cal
I}_1 +4\,Re(g_L\,g^*_R){\cal I}_2\right)-(1 \leftrightarrow 2),
\end{eqnarray}
\begin{equation}
\Delta
Q_V=-\frac{Q_1}{16\pi^2}\left(|g_L|^2+|g_R|^2\right)8\,{\cal
I}_3-(1 \leftrightarrow 2),
\end{equation}
where ${\cal I}_i$ is a parametric integral depending on $m_1$ and
$m_2$. We have omitted a possible color factor, which should be
inserted when appropriate. $(1 \leftrightarrow 2)$ stands for an
additional term in which the interchanges $Q_1 \leftrightarrow
Q_2$ and $m_1 \leftrightarrow m_2$ are to be made. The parametric
integrals read

\begin{eqnarray}
&&{\cal I}_1=\int^1_0dx \int^{1-x}_0 dy\left(\frac{m^2_1\,x+m^2_V
h(x,y)}{{\cal R}} -(1-2x-y)\log{\cal R}\right),\\
&&{\cal I}_2=\int^1_0dx\int^{1-x}_0dy
\frac{m_1\,m_2(1-2x-2y)}{{\cal
R}},\\
&&{\cal I}_3=\int^1_0dx\int^{1-x}_0dy \frac{m^2_V(1-x-y)xy}{{\cal
R}},
\end{eqnarray}
with
\begin{equation}
{\cal R}=m^2_2-(m^2_V+m^2_2-m^2_1)(x+y)+m^2_V(x+y)^2,
\end{equation}

\begin{equation}
h(x,y)= 2y^3+(5x-3)y^2+(1-2x)^2 y+(x-1)x^2.
\end{equation}
As for $A_V$, it is zero provided that $Q_1= Q_2$, {\it i.e.} for
a neutral no self-conjugate vector boson $V^0$. Since $A_V$ is
associated with a coupling of the vector boson to the photon,
which is induced by the covariant derivative, this result reflects
the fact that a neutral particle can couple to the photon only
through the field tensor $F_{\mu \nu}$. Therefore, the amplitude
for the $V^0 V^{0*}\gamma$ coupling is nonzero and free of
ultraviolet divergences.

After some algebra, the parametric integrals can be solved
explicitly:
\begin{equation}
\fl \Delta
\kappa_{V}=\frac{Q_1}{16\pi^2}\left(\left(|g_L|^2+|g_R|^2\right){\cal
A}(x_1,x_2)+Re(g_Lg_R^*) {\cal B}(x_1,x_2)\right)-(1
\leftrightarrow 2),
\end{equation}
\begin{equation}
\Delta Q_{V}=\frac{Q_1}{16\pi^2}\left(|g_L|^2+|g_R|^2\right){\cal
C}(x_1,x_2)-(1 \leftrightarrow 2),
\end{equation}
with $x_i=m_i/m_V$. The ${\cal A}$, ${\cal B}$, and ${\cal C}$
functions are given by
\begin{eqnarray}
{\cal
A}(x,y)&=&-\frac{1}{3}+(x^2-y^2)\left(1-2(x^2-y^2)\right)\nonumber\\
&+&2(x^2-y^2)\left(x^2-y^2)^2-x^2\right)\log\left(\frac{x}{y}\right)\nonumber
\\
&-&(x^2-y^2)\left((x^2-y^2)^3+x^2(y^2-2x^2+1)+y^4\right)\frac{f(x,y)}{\delta
(x,y)},
\end{eqnarray}
\begin{eqnarray}
{\cal
B}(x,y)&=&4\,x\,y\Bigg\{2+\left(1-2(x^2-y^2)\right)\log\left(\frac{x}{y}\right)\nonumber\\
&+&\frac{1}{2}\left(1+2(x^2-y^2)^2-3x^2-y^2\right)\frac{f(x,y)}{\delta
(x,y)}\Bigg\},
\end{eqnarray}

\begin{eqnarray}
{\cal
C}(x,y)&=&\frac{2}{3}\Bigg\{\frac{2}{3}+2(x^2-y^2)^2+y^2-3x^2\nonumber\\
&-&2\left((x^2-y^2)^3-2x^2(x^2-y^2)+x^2\right)\log\left(\frac{x}{y}\right)\nonumber
\\
&&+\big((x^2-y^2)^4-x^2y^2-x^2(3x^4-3x^2+1)\nonumber\\
&-&y^2(y^4+x^2y^2-5x^4)\big)\frac{f(x,y)}{\delta (x,y)}\Bigg\},
\end{eqnarray}
with
\begin{equation}
\delta (x,y)=\sqrt{1-2(x^2+y^2)+(x^2-y^2)^2},
\end{equation}
\begin{equation}
f(x,y)=\log\left(\frac{1-(x^2+y^2)-\delta
(x,y)}{1-(x^2+y^2)+\delta (x,y)}\right).
\end{equation}
Both $\Delta \kappa_{V}$ and $\Delta Q_{V}$ are antisymmetric
under the interchange $1 \leftrightarrow 2$.

\subsection{CP-odd form factors}
The Feynman parameter technique yield
\begin{equation}
\Gamma^{CP-odd}_{\alpha \beta
\mu}=\frac{8\,i\,e}{16\pi^2}Im(g_Lg^*_R)Q_1{\cal
I}_4\,\epsilon_{\alpha \beta \mu
\lambda}q^\lambda+(1\leftrightarrow 2),
\end{equation}
where

\begin{equation}
{\cal I}_{4}=\int^1_0dx\int^{1-x}_0dy\frac{m_1m_2}{{\cal R}}.
\end{equation}
We have used Shouten's identity to eliminate any redundant term.
The form factor $\Delta \widetilde{\kappa}$ can be written as
\begin{equation}
\label{Delta_k_t} \Delta
\widetilde{\kappa}_{V}=\frac{Im(g_Lg^*_R)}{16\pi^2} \,Q_1\,{\cal
F}(x_1,x_2)+(1\leftrightarrow 2),
\end{equation}
where
\begin{equation}
{\cal
F}(x,y)=4\,x\,y\left(\log\left(\frac{x}{y}\right)+\left(1-x^2+y^2\right)\frac{f(x,y)}{2\delta(x,y)}\right).
\end{equation}
This result is in agreement with that obtained in reference
\cite{Burgess} for the case of the $W$ gauge boson in the context
of left-right symmetric models. From equation (\ref{Delta_k_t}) we
can see that $\Delta \widetilde{\kappa}_{V}$ is symmetric under
the interchange $1\leftrightarrow 2$.

Notice that there is no fermion contribution to $\Delta
\widetilde{Q}$ at the one-loop level, which means that the
electric dipole and magnetic quadrupole moments arising from
fermion loops are entirely determined by $\Delta
\widetilde{\kappa}$. This property was first noted in reference
\cite{Burgess} for a charged $W$ boson with left- and right-handed
couplings to fermions. We have shown that this result is valid in
general, regardless the charge of the gauge boson. There follows
that the fermion contribution to $\widetilde{\mu}_{V}$ is directly
proportional to $\widetilde{Q}_{V}$ at the one-loop level in any
renormalizable theory:
\begin{equation}
2\widetilde{\mu}_{V}+m_{V}\widetilde{Q}_{V}=0.
\end{equation}

\subsection{Form factors of a no-self conjugate neutral gauge boson}

One interesting case is that of a no self-conjugate neutral gauge
boson $V^0$. From the above expressions, one can readily obtain
the results for this case once the replacement $Q_1=Q_2=Q$ is
done:

\begin{eqnarray}
\Delta \kappa_{V^0}&=&\frac{Q}{16\pi^2}\left(
\left(|g_L|^2+|g_R|^2\right){\cal A}_0(x_1,x_2)
+Re(g_Lg^*_R){\cal B}_0(x_1,x_2)\right), \\
\Delta Q_{V^0}&=&\frac{Q}{16\pi^2}\left(|g_L|^2+|g_R|^2\right){\cal C}_0(x_1,x_2),\\
\Delta
\widetilde{\kappa}_{V^0}&=&\frac{Q\,Im(g_L\,g^*_R)}{16\pi^2}{\cal
F}_0(x_1,x_2),
\end{eqnarray}
where
\begin{eqnarray}
{\cal
A}_0(x,y)&=&2(x^2-y^2)-2(x^2-y^2)^2\log\left(\frac{x}{y}\right)
\nonumber\\
&+&(x^2-y^2)\left((x^2-y^2)^2-x^2-y^2\right)\frac{f(x,y)}{\delta(x,y)},
\\
{\cal B}_0(x,y)&=&8x y\left(\log\left(\frac{x}{y}\right)-
(x^2-y^2)\frac{f(x,y)}{2\delta(x,y)}\right),  \\
{\cal C}_0(x,y)&=&
\frac{4}{3}\left(2(x^2-y^2)^2-x^2-y^2\right)\log\left(\frac{x}{y}\right)+\frac{8}{3}(y^2-x^2),\nonumber\\
&-&\frac{2}{3}\left(2(x^2-y^2)^3+x^2(1-3x^2)-y^2(1-3y^2)\right)
\frac{f(x,y)}{\delta(x,y)}\\
{\cal F}_0(x,y)&=&\frac{4\,x\,y\,f(x,y)}{\delta(x,y)}.
\end{eqnarray}

It is clear that $\Delta \kappa_{V^0}$ and $\Delta Q_{V^0}$ are
antisymmetric under the interchange $x_1 \leftrightarrow x_2$,
whereas $\Delta \widetilde{\kappa}_{V^0}$ is symmetric. As a
consequence the CP-even form factors of a neutral no
self-conjugate boson vanish when the fermion masses are
degenerate.

\section{General behavior of the fermion contribution to the $V$ form factors}
\label{dis}

\subsection{Nondecoupling effects of heavy fermions}
In this section we will analyze the behavior of the form factors
in the decoupling limit of the fermion masses. Since $\Delta
\kappa_V$ and $\Delta \widetilde{\kappa}_V$ are associated with
Lorentz structures that arise from dimension-four operators, it is
expected that they are sensitive to nondecoupling effects of heavy
physics. On the other hand $\Delta Q_V$ and $\Delta
\widetilde{Q}_V$ cannot be sensitive to this class of effects
since they are associated with a Lorentz structure generated by a
nonrenormalizable dimension-six operator \cite{Inami}. In order to
analyze the decoupling properties of the form factors of the $V$
boson, we will focus on the loop amplitudes ${\cal A}$, ${\cal
B}$, ${\cal C}$, and $\cal F$ rather than in the form factors
themselves. We will thus not make any assumption about the values
of the parameters $g_L$, $g_R$ and the electric charges $Q_1$ and
$Q_2$, which indeed are model dependent. The loop amplitudes are
given in terms of the parametric integrals ${\cal I}_i$ as
follows: ${\cal A}$ depends on ${\cal I}_1$, ${\cal B}$ on ${\cal
I}_2$, ${\cal C}$ on ${\cal I}_3$, and $\cal F$ on ${\cal I}_4$.
From the analysis of these integrals we will be able to draw some
general conclusions. It is thus not necessary to consider those
terms obtained after making the replacement $1\leftrightarrow 2$.

To find whether or not any heavy physics effect is present in the
loop amplitudes and found the sources of nondecoupling effects, it
is convenient to analyze the behavior of the parametric integrals
${\cal I}_i$ in the heavy fermion limit. These integrals are given
in terms of  $m^2_V/{\cal R}$, $m^2_2/{\cal R}$, $m_1\,m_2/{\cal
R}$, and $(1-2x-y)\log{\cal R}$. It is evident that any integral
depending on $m^2_V/{\cal R}$ vanishes as $(m_V/m_i)^2$ when
$m_i\gg m_V$, with $m_i$ either $m_1$ or $m_2$. This is true for
${\cal I}_3$, which means that $\cal C$ is insensitive to the
effects of heavy fermions as it only depends on this integral. Now
consider the term $(1-2x-y)\log{\cal R}$. It can be shown that the
respective integral tends to a nonzero constant value when $m_1\to
\infty$ or $m_2\to \infty$, and vanishes when both $m_1$ and $m_2$
become large. It follows that those integrals depending
logarithmically on ${\cal R}$ are sensitive to heavy fermions when
either $m_1$ or $m_2$ is made large and the other one is kept
fixed. Let us now analyze the behavior of the term $m^2_2/{\cal
R}$, which also contribute to ${\cal I}_1$. When $m_1$ becomes
large and $m_2$ remains small, $m^2_2/{\cal R}\to 0$, but in the
opposite case it tends to a constant value. Of course when both
$m_1$ and $m_2$ become heavy, $m^2_2/{\cal R}$ tend to a nonzero
constant. Therefore, the integral ${\cal I}_1$ is sensitive to
nondecoupling effects when both $m_1$ and $m_2$ are heavy or when
either $m_1$ or $m_2$ is heavy. Finally, let us discuss the
behavior of the term $m_1\,m_2/{\cal R}$, which enters into the
integrals ${\cal I}_2$ and ${\cal I}_4$. When $m_1\gg m_2$ both
${\cal I}_2$ and ${\cal I}_4$ decrease as $m_2/m_1$ and vice
versa. On the other hand, when $m_1\sim m_2\gg m_V$, these
integrals tend to a nonzero constant value. Therefore, ${\cal B}$
and $\cal F$ are sensitive to nondecoupling effects only when the
fermions are degenerate, in contrast with ${\cal A}$, which always
is sensitive to heavy fermions.

\begin{figure}
\centering
\begin{tabular}{cc}
\includegraphics[width=2.0in]{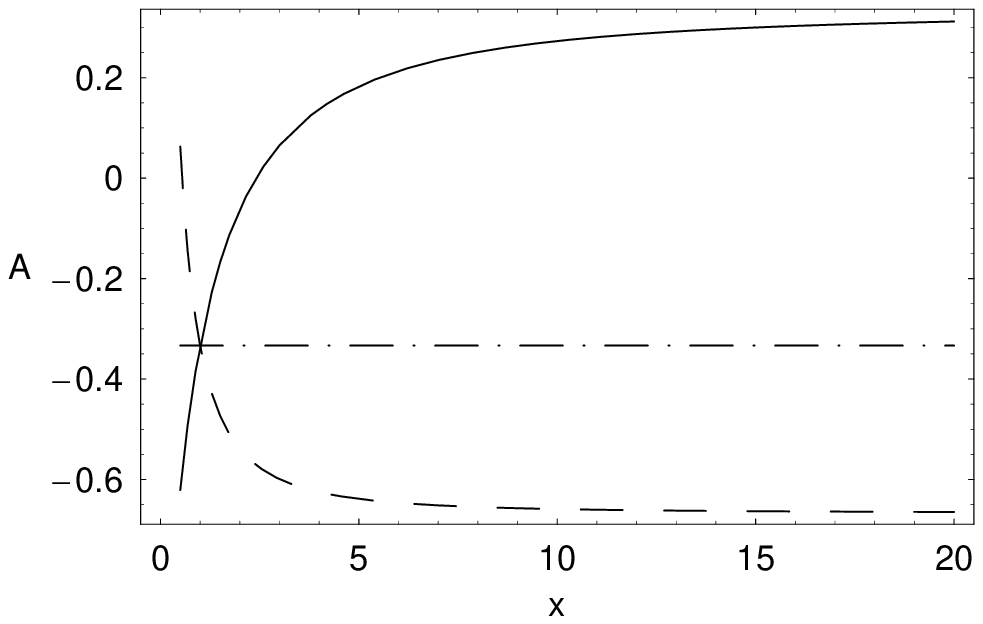}&\includegraphics[width=2.0in]{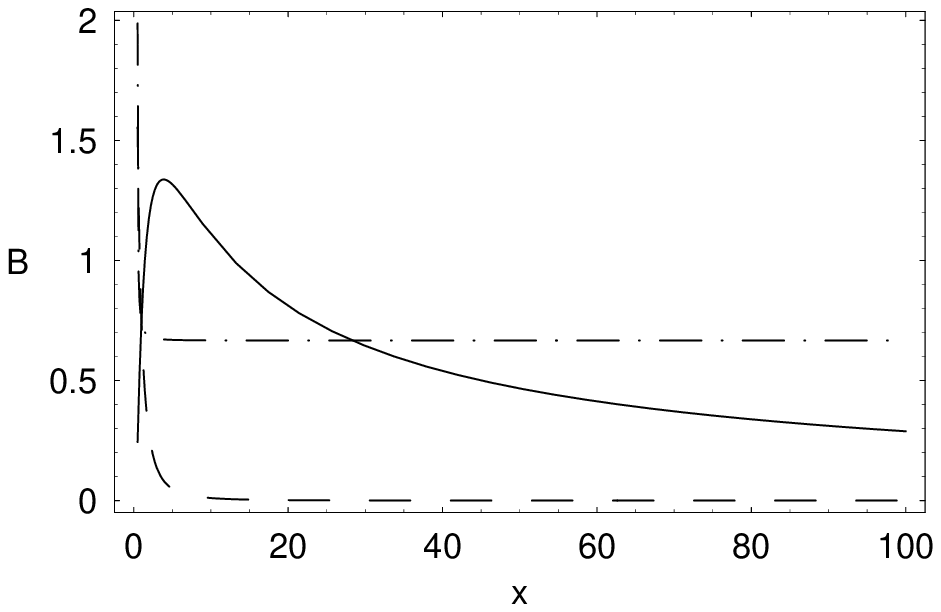}\\
\includegraphics[width=2.0in]{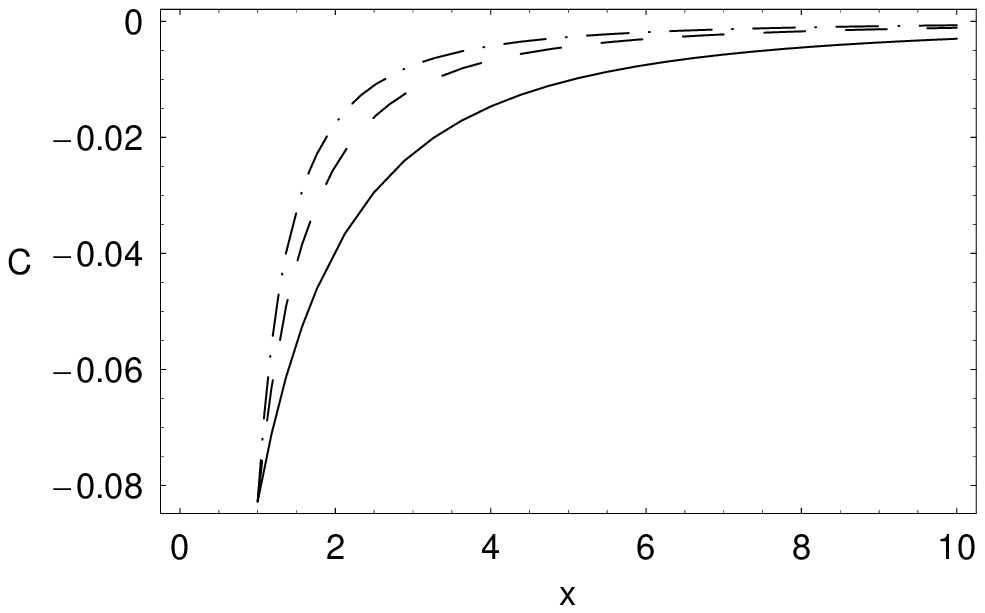}&\includegraphics[width=2.0in]{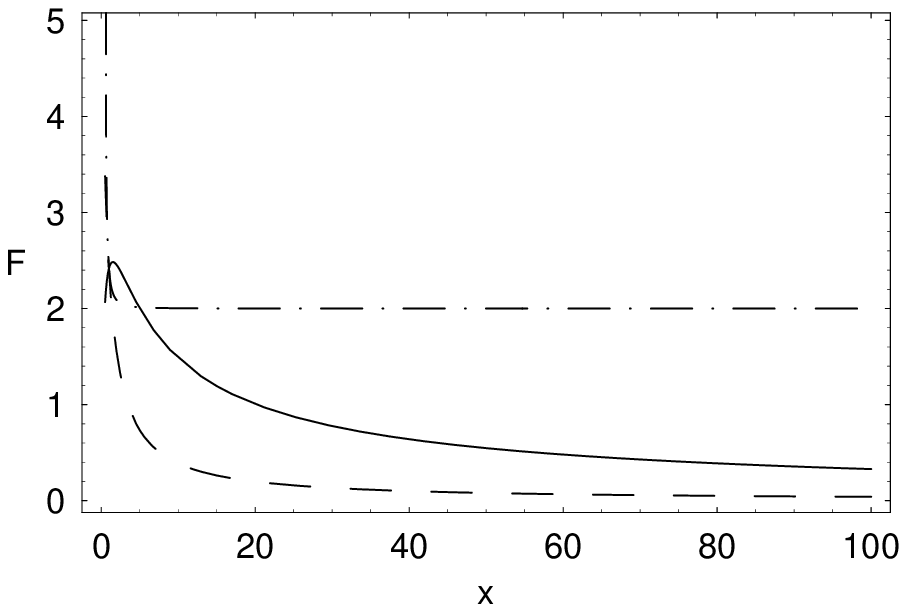}
\end{tabular}
\caption{\label{Func} The fermion-loop amplitude as a function of
$x_1=m_1/m_V$ and $x_2=m_2/m_V$. The curves corresponds to the
following scenarios: $x_1=1$ and $x_2=x$ (solid line), $x_2=1$ and
$x_1=x$ (dashed line), and $x_1=x_2=x$ (dot-dashed line). In order
to show clearly the decoupling properties of the $\cal B$ and
$\cal F$ functions, a larger range for $x$ has been chosen because
these functions decrease slowly with increasing $x$.}
\end{figure}

In summary, $\cal C$ is always insensitive to nondecoupling
effects, $\cal B$ and $\cal F$ are sensitive to heavy fermions
only when the fermion masses are degenerate, and $\cal A$ is
sensitive to nondecoupling effects in both the degenerate and
nondegenerate scenarios. This is illustrated graphically in figure
\ref{Func}. We thus can conclude that while $\Delta \kappa_V$ is
sensitive to heavy fermion effects, $\Delta Q_V$ is always of
decoupling nature. As far as $\Delta \widetilde{\kappa}_V$ is
concerned, it is sensitive to nondecoupling effects provided that
the fermion masses are degenerate. All these results are in
accordance with the decoupling theorem  \cite{Appelquist}, which
establishes that only those Lorentz structures arising from
renormalizable operators can be sensitive to nondecoupling
effects, whereas those structures coming from nonrenormalizable
operators are suppressed by inverse powers of the heavy mass. In
most of the theories, the nondecoupling effects are unobservable
since they are absorbed by renormalization \cite{Appelquist}.

It is worth analyzing separately the specific case of a neutral no
self-conjugate vector boson.  From equations (7) and (8) it is
clear that the CP-even form factors vanish when $Q_1=Q_2$ and
$x_1=x_2$ since they are antisymmetric under this interchange. In
the case that $x_1\ne x_2$, $\Delta \kappa_{V^0}$ is sensitive to
nondecoupling effects when either $x_1\gg x_2$ or $x_2\gg x_1$,
and the same is true for $\Delta \widetilde{\kappa}_V$. As far as
$\Delta Q_{V}$ is concerned, it is of decoupled nature and
vanishes when either $x_1$ or $x_2$ are heavy. The latter form
factor goes to zero rapidly in the heavy fermion limit.

Although the actual size of these form factor is model dependent,
we can infer their order of magnitude from the analysis of the
loop functions. In figure \ref{Func} we show these loop functions
in three scenarios of interest: $x_1=1$ and $x_2=x$, $x_2=1$ and
$x_1=x$, and $x_1=x_2=x$. The first two scenarios corresponds to
the case of a fermion with mass of the order of $m_V$, whereas the
third one corresponds to the case of degenerate fermions. From
these figures we can see that $\cal A$, $\cal B$ and $\cal F$ can
be of the order of $O(1)$, whereas $C$ is one or two orders of
magnitude below. Thus if there is no large cancellations, $\Delta
\kappa$ and $\Delta \widetilde \kappa$ can be at most of the order
of $g^2/(16 \pi^2)$, whereas $\Delta Q$ is one or two orders of
magnitude below. However, $\Delta \widetilde \kappa$ requires that
both $g_L$ and $g_R$ are nonzero and may be further suppressed.

\subsection{Degeneracy and nondegeneracy of the fermion masses}

Let us now analyze the static quantities of the $V$ boson in two
scenarios of interest.  In particular, we will obtain explicit
expressions for the form factors in the massless fermion and heavy
fermion limits. In the first scenario we assume that there is
degeneracy of the fermion masses and in the second case we assume
$x_2=0$ and $x_2$ arbitrary, which implies that one fermion is
massless or much lighter than the other one. We rewrite the static
quantities of the $V$ boson as

\begin{equation}
\Delta
\kappa_V=\frac{1}{16\,\pi^2}\left(\left(g^2_L+g^2_R\right)\bar{{\cal
A}}(x)+Re(g_L\,g_R^*)\bar{{\cal B}}(x)\right),
\end{equation}
\begin{equation}
\Delta Q_V=\frac{1}{16\,\pi^2}\left(g^2_L+g^2_R\right)\bar{{\cal
C}}(x),
\end{equation}
and
\begin{equation}
 \Delta\widetilde
\kappa_V=\frac{1}{16\,\pi^2}Im(g_L\,g_R^*)\bar{{\cal F}}(x),
\end{equation}

In the scenario with degenerate fermions ($x_1=x_2=x$) we obtain

\begin{eqnarray}
\label{Abar}
\bar{{\cal A}}(x)&=&\frac{1}{3}(Q_2-Q_1), \\
\label{Bbar} \bar{{\cal
B}}(x)&=&8(Q_1-Q_2)x^2\left(1-\sqrt{4x^2-1}
\tan^{-1}\left(\frac{1}{\sqrt{4x^2-1}}\right)\right), \\
\label{Cbar} \bar{{\cal
C}}(x)&=&\frac{4}{9}(Q_2-Q_1)\left(-1+3x^2+\frac{6x^2(1-2x^2)}{\sqrt{4x^2-1}}
\tan^{-1}\left(\frac{1}{\sqrt{4x^2-1}}\right)\right),\\
\label{Fbar} \bar{\cal
F}&=&4(Q_1+Q_2)\frac{x^2}{\sqrt{4x^2-1}}\tan^{-1}\left(\frac{1}{\sqrt{4x^2-1}}\right).
\end{eqnarray}

Some conclusions can be drawn from these expressions. First of
all, $\bar{{\cal A}}$ is constant, which means that $\Delta
\kappa_V$ always receives a contribution from a degenerate fermion
doublet, regardless the value of the fermion mass. Notice also
that $\bar{{\cal B}}$ vanishes for $x=0$, whereas $\bar{{\cal C}}$
takes the value $(4/9)(Q_1-Q_2)$. From here we can recover the
known result for the contribution from massless fermions to the
static quantities of the $W$ boson in the SM \cite{Bardeen}. In
the heavy fermion limit, $\bar{{\cal B}}(x)=(2/3)(Q_1-Q_2)$ and
$\bar{{\cal C}}(x)=0$, which is also in accordance with the
previous discussion. As for the CP-odd form factor $\bar{\cal F}$,
it vanishes when $x=0$ and tends to $2(Q_1+Q_2)$ in the heavy
fermion limit. It means that $\Delta\widetilde \kappa$ is
insensitive to massless fermions but sensitive to a heavy
degenerate fermion doublet. From equations
(\ref{Abar})-(\ref{Fbar}) it is also clear that in this scenario
there are no contributions to the CP-even form factors of a
neutral $V^0$ boson, whereas the CP-odd contribution is nonzero.
This result is independent of the fermion mass.

Let us now consider the scenario with nondegenerate fermions ($x_1
\ne x_2$). Without losing generality we consider $x_2=0$ and
$x_1=x$. We obtain
\begin{eqnarray}
\bar{{\cal
A}}(x)&=&\frac{1}{3}(Q_2-Q_1)(1+6x^4)+(Q_1+Q_2)x^2\nonumber\\&+&2x^4\left((Q_2-Q_1)x^2+Q_1\right)
\log\left(\frac{|x^2-1|}{x^2}\right), \\
\bar{{\cal
C}}(x)&=&-\frac{4}{9}\left(Q_2-Q_1\right)(1+3x^4)-\frac{2}{3}(Q_2+3Q_1)x^2\nonumber\\&-&\frac{4}{3}x^2
\left((Q_2-Q_1)x^4
+2Q_1x^2-Q_1\right)\log\left(\frac{|x^2-1|}{x^2}\right),
\end{eqnarray}
and $\bar{\cal B}=0$. In the heavy-mass limit, $\bar{{\cal
A}}\to-(2\,Q_1+Q_2)/3$ and $\bar{C}\to 0$, which again is in
agreement with the previous discussion. As long as the CP-odd
amplitude is concerned,  there is no contribution to $\Delta
\widetilde{{\kappa}}_V$ in this scenario as it vanishes when
either $x_1=0$ or $x_2=0$.

The respective expressions for a neutral vector boson can be
readily obtained  after the replacement $Q_1=Q_2=Q$ is done:
\begin{eqnarray}
\bar{\cal
A}_0(x)&=&2Qx^2\left(1+x^2\log\left(\frac{|x^2-1|}{x^2}\right)\right),
\\
\bar{\cal
C}_0(x)&=&-\frac{4}{3}Qx^2\left(2+(2x^2-1)\log\left(\frac{|x^2-1|}{x^2}\right)\right).
\end{eqnarray}
In the heavy-mass limit, $\bar{\cal A}_0\to -Q$ and $\bar{\cal
C}_0\to 0$. Of course when $x=0$, the degenerate fermion case is
recovered and there is no contributions to the CP-even form
factors.

\section{Summary} \label{con}
We presented a comprehensive study of the one-loop fermion
contributions to the static electromagnetic properties of a no
self-conjugate vector boson with arbitrary electric charge. A
renormalizable coupling of the $V$ boson to left- and right-handed
fermions was assumed.  Analytical expressions for the respective
form factors were presented in the general case and some scenarios
of interest, which can be used to evaluate the fermion
contribution in the context of any renormalizable theory
predicting this class of interactions. Apart from the two CP-even
form factors, the fermion loop gives rise to only one CP-odd form
factor, namely $\Delta \widetilde{\kappa}_V$. As a consequence,
the contributions to the electric dipole and magnetic quadrupole
moments turn out to be directly proportional to each other. This
result is valid for any no self-conjugate vector boson, regardless
its electric charge. As for the CP-odd form factor $\Delta
\widetilde{\kappa}_V$, it turns out to be proportional to
$Im(g_L\,g^*_R)$, thereby requiring the presence of both left- and
right-handed fermions.

The behavior of the electromagnetic form factors of the $V$ boson
was analyzed in several scenarios. Particular emphasis was given
to the heavy fermion limit, and the consistency with the
decoupling theorem was discussed. As for the CP-even form factors,
it was shown that $\Delta \kappa_V$ is sensitive to heavy
fermions, whereas $\Delta Q_V$ is always of decoupling nature. The
latter decreases rapidly, as $m_i^{-2}$, when the heavier fermion
mass $m_i$ increases, which means that it is only sensitive to
physics effects which are not very far from the $m_V$ scale. The
CP-odd quantity $\Delta \widetilde{\kappa}_V$ is sensitive to
heavy fermions provided that the fermions are degenerate, whereas
in a nondegenerate scenario it decreases as $m_i^{-1}$. It was
also found that for fermion masses near the $m_V$ scale, the loop
amplitudes associated with $\Delta \kappa_V$ and $\Delta
\widetilde{\kappa}_V$ are of the same order of magnitude, but the
one associated with $\Delta Q_V$ is one order of magnitude below.
Since the $\Delta \widetilde{\kappa}_V$ form factor requires the
presence of both left-handed and right-handed fermions, it depends
strongly on the existence of a complex phase, which is expected to
be very small in most of the renormalizable theories.

We would like to emphasize that all the above remarks are valid
regardless the electric charge of the $V$ boson. An interesting
case is that of a neutral no self-conjugate vector boson, for
which explicit analytical expressions were obtained. In
particular, it was found that there is no contribution to the
CP-even form factor of the neutral vector boson when the fermion
masses are degenerate, in contrast to the CP-odd form factor,
which is nonzero in this scenario.

\ack{Support from CONACYT and SNI (M\' exico) is acknowledged.
Partial support from SEP-PROMEP and
VIEP-BUAP is also acknowledged.\\[1cm]}

\end{document}